\documentclass[11pt]{article}

\usepackage{epsfig,color}
\usepackage{amsmath,amssymb}
\usepackage{amssymb}

\renewcommand{\l}{\lambda}
\newcommand{\G}{\Gamma}

\def\pa{\partial}

\renewcommand{\d}{\delta}

\newcommand{\m}{\mu}
\newcommand{\n}{\nu}

\newcommand{\D}{\Delta}

\def\be{\begin{equation}}
\def\ee{\end{equation}}
\def\bea{\begin{eqnarray}}
\def\eea{\end{eqnarray}}
\def\ba{\begin{array}}
\def\ea{\end{array}}
\def\bi{\begin{itemize}}
\def\ei{\end{itemize}}

\begin{document}
\thispagestyle{empty}
\null\vskip-24pt \hfill AEI-2001-035 \vskip-10pt
\hfill LAPTH-841/01 \vskip-10pt
\hfill
IFUM-FT-683 \vskip-10pt \hfill {\tt hep-th/0103230} \vskip0.2truecm
\begin{center}
\vskip 0.2truecm {\Large\bf
Exceptional non-renormalization properties and
OPE analysis of chiral four-point functions in ${\cal
N}=4$ SYM$_4$}\\
\vskip 0.5truecm
{\bf
G. Arutyunov$^{*,**}$\footnote{email:{\tt agleb@aei-potsdam.mpg.de}},
B. Eden$^{\ddagger }$\footnote{email:{\tt burkhard@lapp.in2p3.fr}},
A. C. Petkou$^{\dagger}$ \footnote{email:{\tt Anastasios.Petkou@mi.infn.it}},
E. Sokatchev$^{\ddagger}$ \footnote{email:{\tt sokatche@lapp.in2p3.fr} \\
$^{**}$On leave of absence from Steklov Mathematical Institute,
Gubkin str.8, 117966, Moscow, Russia
}
}\\
\vskip 0.4truecm
$^{*}$
{\it Max-Planck-Institut f\"ur Gravitationsphysik,
Albert-Einstein-Institut, \\
Am M\"uhlenberg 1, D-14476 Golm, Germany}\\
\vskip .2truecm $^{\ddagger}$
{\it Laboratoire
d'Annecy-le-Vieux de Physique Th{\'e}orique\footnote{UMR 5108
associ{\'e}e {\`a} l'Universit{\'e} de Savoie} LAPTH, Chemin de
Bellevue, B.P. 110, F-74941 Annecy-le-Vieux, France} \\
\vskip .2truecm $^{\dagger}${\it Dipartimento di Fisica
dell'Universita di Milano and I.N.F.N. Sezione di Milano,
\\ via Celoria 16,
20133 Milano, Italy}\\
\end{center}

\vskip 1truecm
\Large
\centerline{\bf Abstract}

\normalsize

We show that certain classes of apparently unprotected operators in
${\cal N}=4$ SYM$_4$ do not receive quantum corrections as a consequence
of a  partial non-renormalization theorem for the four-point function of
chiral primary operators.
We develop techniques yielding the asymptotic expansion
of the four-point function of CPOs up to order $O(\l^2)$ and we
perform a detailed OPE analysis. Our results reveal the existence of new
non-renormalized operators of approximate dimension 6.

\newpage
\setcounter{page}{1}

\section{Introduction}
The supersymmetric ${\cal N}=4$ Yang-Mills theory in four dimensions
(SYM$_4$) has recently attracted a lot of attention,
primarily as the prototype example of the AdS/CFT
correspondence \cite{M}-\cite{W}. Additionally, it has been gradually
realized that SYM$_4$ by itself constitutes an interesting
quantum field theoretic model where
some unexpected properties emerge.

Perhaps the most interesting local operators in the theory are the
chiral or analytic operators forming short multiplets of the
superconformal group $SU(2,2|4)$ (see the classification in
\cite{dobPet}). An important class of these \cite{hw1,FZ} can be
written in terms of the ${\cal N}=4$ on-shell superfields $W^i$
($i$ is an index of the irrep {\bf 6} of the R-symmetry group
$SU(4) \sim SO(6)$) as ${\mbox tr}(W^{\{i_1}...W^{i_k\}})$. The
conformal dimensions of short operators as well as their two- and
three-point correlation functions are protected from perturbative
corrections \cite{lmrs}--\cite{psz}, therefore they are
well-suited quantities for tests of the AdS/CFT correspondence.
Other classes of operators in ${\cal N}=4$ SYM$_4$ include
operators dual to massive string modes that decouple at strong
coupling (e.g. the Konishi multiplet) \cite{GKP} and operators
dual to multi-particle supergravity states whose strong coupling
anomalous dimensions are non-zero.

The renormalization properties of gauge invariant operators in
${\cal N}=4 $ SYM$_4$ are to a large extent determined by superconformal
invariance and unitarity \cite{FZ}. A powerful test for the
various predictions regarding the operator algebra is the study
of four-point functions, which encode all the relevant
dynamical information through vacuum operator product expansions (OPEs).
Recently the four-point function of the chiral primary operators (CPOs),
which are the lowest scalar
components of the short superfield ${\mbox tr}(W^{\{i}W^{j\}})$
comprising the Yang-Mills stress-energy tensor multiplet, were computed
in perturbation theory up to two loops (to order $\lambda^2$)
\cite{EHSSW}-\cite{BKRS2}.

On the gravity side the calculation of the four-point function
of CPOs via the AdS/CFT correspondence is highly complicated
because one first has to establish the relevant part
of the supergravity action for scalar fields corresponding
to these CPOs. For the massless dilaton and axion fields
the action is already known \cite{LT}
and with the development of the powerful integration technique
over the AdS space \cite{dhofre} the complete results for the
four-point functions became available \cite{HFMMR}.
However, the dilaton and axion fields correspond to descendants
of CPOs, which rather complicates the corresponding CFT analysis
\cite{HMMR,H}.
With the evaluation of the quartic supergravity couplings
for scalar fields corresponding to CPOs \cite{AFcoupl} the computation
of the four-point functions of the lowest-weight CPOs in the supergravity
approximation has been recently completed in \cite{AF}.

The CPOs do not form a ring structure with respect to the OPE, i.e.
in general their OPE contains
fields acquiring  non-zero anomalous dimensions.
A partial OPE analysis of the four-point functions of CPOs was performed
in \cite{BKRS1,BKRS2,AFP1, AFP2} and the anomalous dimensions of certain
operators were found both at weak and strong coupling.
The results of these papers show agreement with the
general considerations of \cite{FZ} based on superconformal invariance
and unitarity.
However, the careful analysis of
\cite{AFP1,AFP2} led to a surprise:
the OPE of two lowest-weight CPOs contains operators
whose anomalous dimensions
vanish both at weak and at strong coupling,
although they are
apparently not protected by unitarity.
Such an unexpected result indicates the existence of new
non-renormalization theorems
in ${\cal N}=4$ SYM that may be a consequence of the dynamics of
the gauge theory rather than its kinematics.

The superconformal properties of
the ${\cal N}=4$ SYM are accounted for very clearly by formulating the
theory in ${\cal N}=2$ harmonic superspace \cite{hh}. In this formulation
the analogues of the ${\cal N}=1$ chiral matter superfields obey the
constraint of $G$-analyticity while their equations of motion
take the form of $H$-analyticity. In a recent paper
\cite{EPSS} it was shown that superconformal covariance and
the requirements of $G$- and $H$-analyticity
combined with Intriligator's insertion formula \cite{I},
constrain the four-point correlation functions of the lowest-weight CPOs
(a priory given by {\it two} arbitrary functions of conformal variables)
to depend on a {\it single} function $F$, which in addition obeys constraints
from crossing symmetry. This function comprises
all possible quantum corrections (perturbative and instanton) to the
free-field result.

In the present paper we show that non-renormalization of some operators, that
are not in general protected by unitarity restrictions \cite{AFP1,AFP2},
follows from the partial non-renormalization theorem of \cite{EPSS}. A typical
example which we study in some detail is a scalar operator $O^{\bf 20}$ of
conformal dimension 4 transforming in the irrep $\bf 20$ of $SO(6)$. In
free-field theory it can be represented by a ``double trace'' operator
\mbox{$:\mbox{tr}(\phi^i\phi^j)\mbox{tr}(\phi^k\phi^l):$}, where $\phi^i$ are
the six scalars of {\cal N}=4 Yang-Mills and the $SO(6)$-indices are projected
onto the $\bf 20$. In free-field theory this operator saturates the unitarity
bound of the A') series of UIRs in the classification of \cite{FZ} but in an
interacting theory it can in principle acquire an anomalous dimension.

To find implications of the partial
non-renormalization for the OPE
of short operators we use Conformal Partial
Wave Amplitude (CPWA) analysis, a subject
well developed in the past \cite{FGG}-\cite{FP}
and recently revitalized in the context of the AdS/CFT duality
\cite{L,HMMR,HPR,AFP1,AFP2,DO,DO1}.
In a conformally invariant theory the OPE of two scalar
fields is decomposed in terms of conformal blocks of traceless
symmetric tensors. Each of them realizes an irreducible
representation of the conformal group.
The CPWA can be viewed as the contribution of the conformal block
of a tensor field to the conformally covariant four-point function.
For the four-point functions considered here, due to the
universality of the quantum correction function $F$, the
projections onto different irreps of the R-symmetry
group are related to each other. Matching these relations against
the CPWA expansion of the various projections
we are able to demonstrate the absence of quantum corrections to some
operators in the ${\bf 20}$ and the ${\bf 105}$.

Apart from the non-renormalized operators just discussed, there
exist other operators that do not receive quantum corrections \cite{AFP1}.
However, their non-renormalization properties
are encoded in the explicit form of the function $F$
whose non-perturbative expression is currently unavailable.
The function $F$ contains
information about both protected and unprotected operators, but
the latter are mixed in perturbation theory. To solve
for the operator mixing one has two possibilities. Firstly, one may
compute the weak coupling four-point functions
of other fields appearing in the OPE of CPOs and then find and
diagonalize the corresponding mixing matrix. Secondly, one
may exploit the partial knowledge of $F$ in different regimes.
Here we employ the second possibility to trace new non-renormalized operators.

Since the two-loop ($O(\l^2)$) four-point function is
known \cite{ESS,BKRS2}, we can use it to extract the corresponding OPE
expansion.
In view of comparison with a sum of CPWAs of different tensors
it is therefore desirable to represent this function as a
certain series expansion valid in the asymptotic region
of conformal variables where we study the OPE.
We solve this problem by using an analytic regularization
that allows one to reduce the two-loop function
to the function related with a one-loop diagram.
Our approach is different from the one in \cite{BKRS2}.
Using the CPWA analysis of the one-loop, two-loop
and strong coupling four-point functions of CPOs we then demonstrate
the vanishing of the anomalous dimension for the scalar
of naive dimension 6 transforming in the {\bf 20} of $SO(6)$.

The paper is organized as follows. In Section 2 we start by recalling
the structure of the four-point functions of the lowest-weight CPOs
and describe the partial non-renormalization theorem of \cite{EPSS}.
Employing CPWA analysis we show that the absence of quantum corrections
to $O^{\bf 20}$ and to the rank-$2k$ tensors of dimension $4+2k$
is a direct consequence of the partial non-renormalization
of the four-point functions of CPOs.
In Section 3 we derive a series representation (with logs)
for the two-loop four-point function of CPOs suitable for the study
of the OPE. Some results of the CPWA analysis relevant
for further non-renormalization issues are
presented in Section 4. The technical details
are relegated to two Appendices.

\section{Partial non-renormalization of the four-point function of CPOs}
\setcounter{equation}{0}
In the notation of \cite{AFP1,AFP2}, the four-point function of the lowest
dimension canonically normalized CPOs  is
\begin{eqnarray}
\label{4ptgen} \langle O^{I_1}(x_1)
O^{I_2}(x_2)O^{I_3}(x_3)O^{I_4}(x_4)\rangle &=& a_1(x) \,
\d^{I_1I_2}\d^{I_3I_4} + a_2(x) \, \d^{I_1I_3}\d^{I_2I_4}
\nonumber
\\&&\hspace{-6cm} + a_3(x) \, \d^{I_1I_4}\d^{I_2I_3}
+ b_2(x) \, C^{I_1I_2I_3I_4} + b_1(x) \, C^{I_1I_3I_2I_4} + b_3(x)
\, C^{I_1I_3I_4I_2} \, ,
\end{eqnarray}
where $I_1,\ldots ,I_4=1,2,..,20$ are indices of the irrep ${\bf
20}$ of $SO(6)$ and the various $C$-tensors in (\ref{4ptgen}) were
defined in \cite{AFP1}. Here $a_i$ and $b_i$ are given by simple
propagator factors times functions of the two biharmonic ratios
\begin{equation}
\label{uvY}
u=\frac{x_{12}^2x_{34}^2}{x_{13}^2x_{24}^2}
\,,\,\,\,\,v=\frac{x_{12}^2x_{34}^2}{x_{14}^2x_{23}^2}\, .
\end{equation}
In the sequel we will also use the variable $Y=1-\frac{v}{u}$.

The Bose symmetry (equivalently crossing symmetry) of the four-point
function (\ref{4ptgen}) implies that only one of the $a_i$'s and
one of the $b_i$'s are independent. A further restriction on the
structure of the four-point function is imposed by the partial
non-renormalization theorem of \cite{EPSS} which states that all
six coefficient functions in (\ref{4ptgen}) are expressed in terms
of {\it one and only one} arbitrary function of two variables
$F(v,Y)$:
\begin{eqnarray}
\label{a1}
a_1 &=& \frac{1}{x_{12}^4x_{34}^4}\left[
1 +u{ F}(v,Y)\right]\\
\label{a2}
a_2 &=& \frac{1}{x_{12}^4x_{34}^4}\left[
u^2 +u^2\,{ F}(v,Y)\right]\\
\label{a3}
a_3 &=& \frac{1}{x_{12}^4x_{34}^4}\left[
v^2 +vu\, {F}(v,Y)\right]\\
\label{b1}
b_1 &=& \frac{1}{x_{12}^4x_{34}^4}\left[\frac{4}{N^2}\,vu+ (vu^2 -u^2
-vu){ F}(v,Y)\right] \\
\label{b2}
b_2 &=& \frac{1}{x_{12}^4x_{34}^4}\left[\frac{4}{N^2}\,v+ (v -vu
-u) { F}(v,Y)\right] \\
\label{b3}
b_3 &=& \frac{1}{x_{12}^4x_{34}^4}\left[
\frac{4}{N^2}\,u+ (\frac{u^2}{v} -u^2
-u){ F} (v,Y)\right] \, .
\end{eqnarray}
Here the $F$-independent terms correspond to the disconnected ($a_i$)
and connected ($b_i$) parts in the free amplitude (Born
approximation). The function ${F}(v,Y)\equiv {\cal F}(u,u/v)$
encodes all {\it quantum corrections} and obeys the crossing
symmetry relations \cite{EPSS}:
\begin{eqnarray}
\label{trans}
{\cal F}(u,u/v)={\cal F}(u/v,u)=\frac{v}{u}{\cal F}(v,v/u) \, .
\end{eqnarray}

Our prime interest will be to understand the implications of this
partial non-renormaliza\-tion theorem for the OPE of chiral
operators. To this end we will discuss the OPE for $x^2_{12}\,
,x^2_{34}\to 0$, or equivalently $u\, , v\, ,Y\to 0$.

The product of two CPOs $O^{I_1}(x_1)O^{I_2}(x_2)$ decomposes
under the $R$-symmetry group $SU(4)$ as
\begin{equation}\label{irreps}
  {\bf 20}\times {\bf 20}={\bf 1}+{\bf 20}+{\bf 105}+{\bf 84}+{\bf 15}+{\bf 175} \, .
\end{equation}
To label the different operators appearing in the operator product expansion we
use the notation $O_{\D,l}^{\bf irrep}$, where $\D$ describes the free-field
conformal dimension of the operator, $l$ is its Lorentz spin and {\bf irrep}
denotes the corresponding representation of $SU(4)$.

By analyzing the four-point function of chiral operators at strong
coupling \cite{AF} it was found in \cite{AFP1} that there exist an
operator $O^{\bf 20}_{4,0}$ and a tower of rank-$2k$ tensors
$O^{\bf 105}_{4+2k,2k}$ which do not acquire anomalous dimension.
In \cite{AFP2} the same phenomenon was observed at the one-loop
($O(\lambda)$) level. These operators do not belong to short
superconformal representations and thus the standard protection
mechanism \cite{FZ} does not apply to them.
The absence of quantum corrections should be interpreted as a
dynamical rather then a kinematical effect.

In this section we demonstrate these new non-renormalization properties
without making use of perturbative arguments. They are, in fact, a
simple consequence of the general non-perturbative form
(\ref{a1})-(\ref{b3}) of the amplitude. The method we use to
extract information about the content of the operator algebra is
Conformal Partial Wave Amplitude (CPWA) analysis \cite{FGG}-\cite{FP}.

The correlator (\ref{4ptgen}) (or any of its projections on the
irreps (\ref{irreps})) can be written as an expansion of the type
\begin{equation}\label{exptype}
  \langle O(1)O(2)O(3)O(4) \rangle =\sum_{\Delta, l} a_{\Delta, l}{\cal H}_{\Delta,
  l}(x_{1,2,3,4})\; .
\end{equation}
Here ${\cal H}_{\Delta, l}(x_{1,2,3,4})$ denotes the CPWA for the exchange of
a symmetric traceless tensor of rank $l$ and of
(possibly anomalous) dimension $\Delta$. The coefficients $a_{\Delta, l}$ are
to be found by matching the explicit form of the left-hand side of eq.
(\ref{exptype}) to that of the CPWAs. The latter were obtained in \cite{HPR}
and are given in Appendix A, equation (\ref{cpwa}), in terms of the variables
$v,Y$ suitable for the study of the OPE in the direct channel. Here we only
list some basic facts about these CPWAs needed for our argument.

Let us split the dimension of the exchanged operator $\Delta =
\Delta_0 + h$, where $\Delta_0$ is an integer and $-1\leq h<1$.
Then the CPWA is a double series of the type
\begin{equation}
\label{gcpwa} {\cal H}_{\Delta, l} = \frac{1}{x_{12}^4 \, x_{34}^4} \,
v^{\frac{h}{2}} \, \sum_{n,m=0}^\infty c^{\Delta, l}_{nm} v^n Y^m \, .
\end{equation}
Note that the factor $v^{\frac{h}{2}}$ is a fractional power of $v$, which will
allow us to treat CPWAs with different $h$ as functionally independent. As we
show in Appendix A all monomials in this expansion obey
\begin{equation}\label{bound}
T \equiv 2n+m \geq \Delta_0 \, .
\end{equation}
The ``ordering parameter" $T$ proves very helpful when comparing power
expansions of the type (\ref{gcpwa}).

The terms in the series (\ref{gcpwa}) with $T_{\rm min} =
\Delta_0$ are of the form (see Appendix A, (\ref{cpwa}))
$v^{\frac{1}{2}(\Delta_0-l)} Y^l$,
$v^{\frac{1}{2}(\Delta_0-(l-2))} Y^{(l-2)}, \dots$ down to
$v^{\frac{1}{2}(\Delta_0-1)} Y$ or $v^{\frac{1}{2}\Delta_0}$
depending on whether $l$ is even or odd. This means that we can
choose $\Delta_0$ even(odd) for even(odd) spins. Further, from the
unitarity bound $\Delta \geq 1$ (if $l=0$) we deduce that the
entire range of scalar dimensions can be covered choosing
$\Delta_0=2,4,6,\ldots$ and $-1 \leq h<1$. If $l>0$ the unitarity
bound becomes $\Delta \geq 2+l$, so we start at $\Delta_0 = 2+l$
(restricting $0\leq h <1$) and further $\Delta_0= 4+l, 6+l,
\ldots$ with $-1 \leq h<1$.

The main question we are addressing here concerns the exchange operator $O^{\bf
20}_{4,0}$ for which $\Delta_0=4$. According to eq. (\ref{bound}), the
corresponding CPWA has $T_{\rm min} =4$. In order to find out whether such a
CPWA can appear in a given projection of the amplitude, we have to
consider all the CPWAs with $T_{\rm min} \leq 4$.
%
%
%
Within a class of equal fractional power $v^{(h/2)}$ these CPWAs
are\footnote{We do not specify the normalization factors of the
CPWAs, it is just assumed that they are non-singular for the range
of dimensions and spins under consideration.}
\begin{itemize}
\item{$\Delta_0=2 \; \;$ scalar: $v + \ldots$}
\item{$\Delta_0=3 \; \;$ vector: $v Y + \ldots$}
\item{$\Delta_0=4 \; \;$ scalar: $v^2 + \ldots$ \, \ \ rank 2 tensor: $v^2 - vY^2+\ldots$}
\end{itemize}
where only the terms with $T_{\rm min}$ are shown.

Let us now try to match a sum of such CPWAs with the four-point
amplitude in the form (\ref{4ptgen}), (\ref{a1})-(\ref{b3})
predicted by the partial non-renormalization theorem of
\cite{EPSS}. Since we are only interested in anomalous dimensions
which come from the quantum ($F$) terms in (\ref{a1})-(\ref{b3}),
we can drop the Born terms. Accordingly, when expanding the
quantum terms we can neglect CPWAs with integer dimension. We
begin by projecting the amplitude (\ref{4ptgen}) onto the various
$SU(4)$ irreps (\ref{irreps}):
\begin{itemize}
\item{Projection on the {\bf 1}:
\begin{equation}
- \frac{1}{x_{12}^4 \, x_{34}^4} \, (20 - 20 Y-\frac{16}{3} v +\frac{10}{3} Y^2
+\frac{8}{3} vY + \frac{1}{3}v^2) \, \Phi(v,Y)
\end{equation}}
\item{Projection on the {\bf 15}:
\begin{equation}
- \frac{1}{x_{12}^4 \, x_{34}^4} \,  (-4Y + 2 Y^2 +vY) \, \Phi(v,Y)
\end{equation}}
\item{Projection on the {\bf 20}:
\begin{equation}
\frac{1}{x_{12}^4 \, x_{34}^4} \, (-\frac{5}{3}v  +\frac{5}{3} Y^2+\frac{5}{6} v Y + \frac{1}{6}v^2 ) \, \Phi(v,Y)
\end{equation}}
\item{Projection on the {\bf 84}:
\begin{equation}
- \frac{1}{x_{12}^4 \, x_{34}^4} \, (-3v +\frac{3}{2}vY +\frac{1}{2}v^2) \,
\Phi(v,Y)
\end{equation}}
\item{Projection on the {\bf 105}:
\begin{equation}
\frac{1}{x_{12}^4 \, x_{34}^4} \, v^2 \, \Phi(v,Y)
\end{equation}}
\item{Projection on the {\bf 175}:
\begin{equation}
\frac{1}{x_{12}^4 \, x_{34}^4} \, v Y \, \Phi(v,Y)
\end{equation}}
\end{itemize}
where we have set
$$\Phi(v,Y) = \frac{v\, F(v,Y)}{(1-Y)^2}\;.$$
Note that the polynomial prefactors have been $T$-ordered.

Consider the projections on the singlet and on the {\bf 20}. Both of them are
supposed to have CPWA expansions of the type (\ref{exptype}):
\begin{equation}\label{pr1}
- \frac{1}{x_{12}^4 \, x_{34}^4} \, (20 - 20 Y-\frac{16}{3} v +\frac{10}{3} Y^2
+\frac{8}{3} vY + \frac{1}{3}v^2) \, \Phi(v,Y) = \sum_{\Delta,l} a^{\bf
1}_{\Delta,l} {\cal H}_{\Delta,l}
\end{equation}
and
\begin{equation}\label{pr20}
\frac{1}{x_{12}^4 \, x_{34}^4} \, (-\frac{5}{3}v  +\frac{5}{3} Y^2+\frac{5}{6} v
Y + \frac{1}{6}v^2 ) \, \Phi(v,Y) = \sum_{\Delta,l} a^{\bf 20}_{\Delta,l} {\cal
H}_{\Delta,l}
\end{equation}
Since the function $\Phi(v,Y)$ is the same in both of these equations, we can
eliminate it and obtain the consistency condition
\begin{equation}\label{consis}
  (\frac{5}{3}v  -\frac{5}{3} Y^2-\frac{5}{6} v Y
- \frac{1}{6}v^2 )\, \sum_{\Delta,l} a^{\bf 1}_{\Delta,l} {\cal H}_{\Delta,l} =
(20 - 20 Y-\frac{16}{3} v +\frac{10}{3} Y^2 +\frac{8}{3} vY + \frac{1}{3}v^2) \,
\sum_{\Delta,l} a^{\bf 20}_{\Delta,l} {\cal H}_{\Delta,l}
\end{equation}
Recall the form (\ref{gcpwa}) of the CPWA, which contains a term
$v^{h/2}$. Different fractional powers of $v$ are functionally
independent, hence the last equation splits into classes of
different $h$. It is enough to investigate the problem for a given
$h$.

We want to know whether the CPWA ${\cal H}_{4+h,0}$, corresponding
to an anomalous dimension ($h\neq 0$) for the operator $O^{\bf
20}_{4,0}$ can appear in the right-hand side of (\ref{consis}).
Let us first assume that $h>0$. This CPWA has $T_{\rm min} = 4$,
therefore we can keep only terms with $T\leq 4$ on both sides of
(\ref{consis}). In the left-hand side we have a polynomial with
$T\geq 2$, so we need only keep the lowest CPWA ${\cal H}_{2+h,0}
\sim v +\ldots$ ($T\geq 2$). In the right-hand side the polynomial
has $T\geq 0$, so we should include several CPWAs:
\begin{equation}\label{eqn}
  (\frac{5}{3}v  -\frac{5}{3} Y^2+ \ldots )\, [a^{\bf 1}_{2+h,0}(v+\ldots) +
  \ldots] =
\end{equation}
$$(20 +\ldots)\left[ a^{\bf 20}_{2+h,0}(v+\ldots)
  + a^{\bf 20}_{3+h,1}(v Y+\ldots)
   + a^{\bf 20}_{4+h,0}(v^2+\ldots)
   + a^{\bf 20}_{4+h,2}(v^2 - vY^2+\ldots) +\ldots    \right]\, .$$
Clearly, the left-hand side has $T\geq 4$, so the first two terms in the
right-hand side with $T<4$ have no match. The crucial point now is that the
polynomial $v^2 - vY^2$ in the left-hand side exactly matches the tensor term
in the right-hand side. Therefore we conclude that
\begin{equation}\label{res}
  a^{\bf 20}_{2+h,0} = a^{\bf 20}_{3+h,1} = a^{\bf 20}_{4+h,0} =0\, .
\end{equation}

It remains to consider the case when $h<0$. In this case the unitarity bound
prevents the CPWAs ${\cal H}_{3+h,1}$ and ${\cal H}_{4+h,2}$ from occurring in
the right-hand side of eq. (\ref{eqn}). Thus, up to order $T=4$ there is no
possible match and this case has to be ruled out.

The vanishing of a coefficient for a CPWA means that there is no
operator with anomalous part of the dimension $h$, for any given value of $h$.
In other words, the scalars of dimension 2,4 and the vector at
dimension 3 remain non-renormalized. The operator $O_{2,0}^{\bf 20}$ is
itself a CPO belonging to a short multiplet of $SU(4)$ and its
non-renormalization is well-known. The absence of a vector in this channel can
be explained by parity. As to $O^{\bf 20}_{4,0}$, we now see that its
non-renormalization is a consequence of the particular structure of the four-point
function dictated by the superconformal invariance and the dynamics of
${\cal N}=4$ SYM$_4$.

The tensor of approximate dimension 4 can be interpreted as the
operator ${\cal K}_{4,2}$ from the Konishi multiplet.


Let us briefly comment on
the irrep {\bf 105}. Here the polynomial factor is $v^3$. We have
pointed out above, that the lowest order of the CPWA of an
operator of free-field dimension $\Delta_0$ and spin $l$ contains
a term $v^{(\Delta_0-l)/2} Y^l$. It follows that any operator in
the {\bf 105} receiving quantum corrections has $\Delta_0-l \geq
6$, yielding a tower of non-renormalized operators $O_{4+2k,2k}$.

Obviously, there should exist other operators which do not receive
quantum corrections; for example descendants of the operators
discussed above. However, at present we do not see an easy way of
unraveling their non-renormalization properties on the general
grounds of the representation (\ref{a1})-(\ref{b3}). In the next
section we show that some other non-renormalized operators exist
and they can be traced by using the knowledge of the function
$F(v,Y)$ in different regimes.

\section{Series representation of the conformal four-point functions}
\setcounter{equation}{0}
In perturbation theory the function ${ F}(v,Y)$ assumes the form of a
series as
\begin{eqnarray}
\label{expan}
{ F}(v,Y)=\frac{1}{N^2}(\tilde{\lambda}{ F}^{(1)}(v,Y)
+\tilde{\lambda}^2{ F}^{(2)}(v,Y)+\ldots)
+{\cal O}\left(\frac{1}{N^4}\right) \, ,
\end{eqnarray}
where $\tilde{\lambda}=\frac{g_{YM}^2N}{(2\pi)^2}$ is the t'Hooft coupling.
In the following we study only the leading terms in $1/N^2$.

The first two terms in the expansion (\ref{expan}) were computed in
\cite{EHSSW}-\cite{ESS}
by using the ${\cal N}=2$ harmonic superspace technique and in
\cite{BKRS1}, \cite{BKRS2}
by means of the ${\cal N}=1$ superspace formalism.
They are given by
\begin{eqnarray}
\label{F1}
{ F}^{(1)}(v,Y)=-2\frac{v}{u}\Phi^{(1)}\left(v,\frac{v}{u}\right) \, .
\end{eqnarray}
and
\begin{eqnarray}
\label{F2}
{ F}^{(2)}(v,Y)&=&
\frac{1}{u}\Phi^{(2)}\left(\frac{1}{v},\frac{1}{u}\right)+
\Phi^{(2)}\left(\frac{u}{v},u\right)
+\frac{v}{u}\Phi^{(2)}\left(\frac{v}{u},v\right)\\
\nonumber
&+&\frac{v}{4u^2}(u+v+uv)
\left(\Phi^{(1)}\left(v,\frac{v}{u}\right)
\right)^2 \, .
\end{eqnarray}
Here the functions $\Phi^{(1,2)}$ admit representations in terms of
the one- and two-loop
box integrals respectively, and they are the first two elements of an infinite
series of conformally covariant ``multi-ladder'' functions introduced
in \cite{UD,B}.

The symmetry properties of the function $\Phi^{(1)}$ are
\begin{eqnarray}
\label{sym}
\Phi^{(1)}\left(u,\frac{u}{v}\right)=\frac{v}{u}
\Phi^{(1)}\left(v,\frac{v}{u}\right),~~~~~
\Phi^{(1)}\left(v,\frac{v}{u}\right)=\frac{1}{v}
\Phi^{(1)}\left(\frac{1}{v},\frac{1}{u}\right) \, .
\end{eqnarray}
One can easily see that by  virtue of (\ref{sym}) the  functions
(\ref{F1}) and (\ref{F2})
obey the symmetry relations  (\ref{trans}).

Since the two-loop correlation function
admits a representation in terms of the two integrals $\Phi^{(1)}$ and $\Phi^{(2)}$, each of them being
covariant under conformal mappings,
it is tempting to suggest that higher loop
correlation functions can be as well represented as certain polynomials
of all possible multi-ladder integrals that can be composed from
field propagators at this level.
To study then the OPE we require the behavior of the
correlation functions in the asymptotic region, where, say, $x_{12}^2\sim 0$
and $x_{34}^2\sim 0$. Furthermore, to develop an efficient
technique for constructing the
field algebra at higher loops we face the difficult problem of
finding an asymptotic expansion of these integrals in terms of
conformally invariant variables valid in the relevant  asymptotic region.
Here we demonstrate that this problem may be overcome by
using the method of analytic regularization \cite{UD} of the
$L$-loop ladder diagram $\Phi^{(L)}$ that allows one to find the latter
in terms of the sum of diagrams related to $\Phi^{(L-1)}$.
Applying this procedure recurrently one will be subsequently left
with a function related to a one-loop diagram.

For the sake of clarity we consider here only the case of the
function $\Phi^{(2)}$ for which
we obtain a series representation (with logs)
in terms of conformal variables $v$ and $Y$.
In the next section this representation will be used to verify
some predictions about the structure of the field algebra
of chiral operators at two loops.
Below we often use notation $y=1-Y=v/u$.

Following \cite{UD} we introduce a function
\begin{eqnarray}
\label{PHId}
\Phi(v,y|\d)=\int \frac{d\l ds}{(2\pi i)^2}\G(-\l)\G(-s)\G(-\l-\d)\G(-s-\d)
\G^2(1+\l+s+\d)v^{\l}y^{s} \, .
\end{eqnarray}
The integration contours run sufficiently close to the imaginary axis
to separate the ascendant and descendent sets of poles. The $s$-integral is convergent
for $|y|<1$ and $|\mbox{arg}\, y|<\pi$.
Using this integral representation
one notices that the function $\Phi(v,y|\d)$ is a particular example
of a general family of $D_{\D_1\D_2\D_3\D_4}$-functions
(see A1)\footnote{The $D$-functions we use here coincide
with the $\bar{D}$-functions (without normalization factor) introduced in \cite{AFP1}.}
which describe contribution of the scalar AdS graphs to
the four-point function \cite{AFP1} of chiral operators computed in
AdS$_5$ supergravity.
Precisely one has the following relation:
\begin{eqnarray}
\label{A}
\Phi\left(v,\frac{v}{u}|\d\right)=D_{1-\d,1,1,1+\d}(v,Y).
\end{eqnarray}
Representation of this type is rather useful since it allows one to
establish a relation between $\Phi$
considered as a function of the conformal variables
in the crossed channels and $D$-functions as functions of
$v,Y$ which parametrize the direct channel.
In the Appendix B we show that the following formulae are valid:
\begin{eqnarray}
\label{ac}
\Phi\left(u,\frac{u}{v}|\d\right)=\left(\frac{v}{u}\right)^{1+\d}
D_{1-\d,1,1+\d,1}(v,Y)\, , ~~~~
\Phi\left(\frac{1}{v},\frac{1}{u}|\d\right)=v^{1+\d}D_{1-\d,1+\d,1,1}(v,Y)\, .
\end{eqnarray}
It is worth pointing out that the sum of parameters $\D_i$ of the
$D$-functions we meet here is equal to four, which is the
dimension of a space-time. This merely reflects the fact that
in our situation $D$-functions coincide with the well-known star-integrals
(the ``box'' diagram in momentum space).
 Evaluating the integral (\ref{PHId}) one gets the following formula in terms
double series in $v,Y$ variables:
\begin{eqnarray}
\nonumber
\Phi(v,Y|\d)&=&\sum_{m,n=0}^{\infty}\frac{Y^m}{m!}\frac{v^n}{(n!)^2}
\Biggl[ \frac{\G(1+\d)\G(-\d)}{\G(1+n+\d)}\frac{\G^2(1+n)\G^2(1+n+m+\d)}{\G(2+2n+m+\d)}
\\
\label{Phids}
&+&v^{-\d}\frac{\G(1-\d)\G(\d)}{\G(1+n-\d)}\frac{\G^2(1+n-\d)\G^2(1+n+m)}{\G(2+2n+m-\d)}
\Biggr] \, ,
\end{eqnarray}
which converges in a neighborhood of $v=0$, $Y=0$.
In the limiting case $\d=0$ one recovers from (\ref{PHId}) the
Mellin-Barnes integral
for $\Phi^{(1)}(v,Y)=\Phi(v,Y|0)$. Taking the limit $\d\to 0$ in (\ref{Phids})
produces the following asymptotic expansion for $\Phi^{(1)}(v,Y)$:
\begin{eqnarray}
\label{Phi1}
\Phi^{(1)}(v,Y)&=&\sum_{n,m=0}^{\infty}\frac{v^nY^m}{(n!)^2m!}
\frac{\G^2(1+n)\G^2(1+n+m)}{\G(2+2n+m)} \nonumber\\
&\times& [-\ln v+2\psi(2+2n+m)-2\psi(1+n+m)] \, .
\end{eqnarray}
This representation was extensively used in \cite{AFP2} to
study the OPE of chiral operators at one loop.

\begin{figure}[htb]
\begin{center}
\input{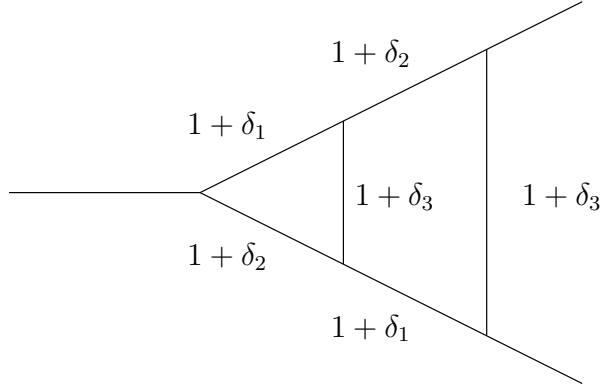}
\end{center}
\caption{Regularized ladder diagram related to the function $\Phi^{(2)}(v,y)$.}
\end{figure}

The idea of \cite{UD} to compute the integral $\Phi^{(2)}(v,y)$ is to introduce
a special analytic regularization of the corresponding two-loop ladder diagram and use
the ``uniqueness'' method to reduce it to the function $\Phi(v,y|\d)$
naturally related to a one-loop ladder diagram. The analytic regularization in question
consists in replacing the powers in denominators by $1+\d_i$ obeying
a condition $\d_1+\d_2+\d_3=0$. After the computation a limit $\d_i=0$ is applied.
In this way one finds the following formula \footnote{The appearance of an additional $\pi^2$ term
in comparison with \cite{UD} is related to the particular series representation for
$\Phi(x,y|\d)$ that we use.}
\begin{eqnarray}
\label{rep}
\Phi^{(2)}(v,y)=\frac{1}{2}\Biggl[
3\pa_{\d}^2\Phi(v,y|\d)-(\ln^2v+\ln v\ln y+\ln^2y+\pi^2)\Phi^{(1)}(v,y)
\Biggr]\, ,
\end{eqnarray}
where the derivative is evaluated at $\d=0$.

With the help of formulae (\ref{ac}) we can also find representations
for function $\Phi^{(2)}$ depending this time on the variables
describing the crossed channels:
\begin{eqnarray}
\label{cr1}
\Phi^{(2)}\left(u,\frac{u}{v}\right)=\frac{y}{2}
\Biggl[3\pa_{\d}^2(y^{\d}D_{1-\d,1,1+\d,1})
-(\ln^2v-3\ln v\ln y+3\ln^2y+\pi^2)\Phi^{(1)}(v,y)
\Biggr]
\end{eqnarray}
and
\begin{eqnarray}
\label{cr2}
\Phi^{(2)}\left(\frac{1}{v},\frac{1}{u}\right)=\frac{v}{2}
\Biggl[3\pa_{\d}^2(v^{\d}D_{1-\d,1+\d,1,1})
-(3\ln^2v-3\ln v\ln y+\ln^2y+\pi^2)\Phi^{(1)}(v,y)
\Biggr] \, .
\end{eqnarray}
One may further simplify the latter expressions by using the fact that
the first derivative of the function $\Phi(v,y|\d)$ computed at $\d=0$
is not independent but can be rather expressed via $\Phi^{(1)}(v,y)$.
To compute the derivatives it is convenient to use the Mellin-Barnes representation for
$\Phi(v,y|\d)$.  Indeed, from (\ref{PHId}),
one can see that under the following shift of integration variables
$\l\to \l-\d/2$ and $s\to s-\d/2$ the function $\Phi(v,y|\d)$ acquires a form
\begin{eqnarray}
\nonumber
\Phi(v,y|\d)&=&(vy)^{-\d/2}
\int \frac{d\l ds}{(2\pi i)^2}
\G(-\l+\d/2)\G(-s+\d/2)\G(-\l-\d/2)\G(-s-\d/2)
\\
\label{MB1}
&\times &\G^2(1+\l+s)v^{\l}y^{s} \, .
\end{eqnarray}
Clearly, viewed as a series in the $\d$-variable,
the integrand does not have a linear term.
This fact allows one to derive an identity \cite{UD}:
\begin{eqnarray}
\pa_{\d}\Phi^{(1)}(v,y)=-\frac{1}{2}\ln vy~\Phi^{(1)}(v,y) \, .
\end{eqnarray}
Similarly, by using the corresponding Mellin-Barnes representations
for the remaining  $D$-functions in (\ref{ac})
(see Appendix B) we obtain the formulae
\begin{eqnarray}
\label{fd}
\pa_{\d}D_{1-\d,1,1+\d,1}=-\frac{1}{2}\ln v \, \Phi^{(1)}\,, ~~~~
\pa_{\d}D_{1-\d,1+\d,1,1}=-\frac{1}{2}\ln y\, \Phi^{(1)} ,
\end{eqnarray}
where the derivatives are taken at $\d=0$ and we omit the arguments.

Now performing the differentiation in (\ref{cr1}), (\ref{cr2})
and using expressions (\ref{fd}) we arrive at
\begin{eqnarray}
\Phi^{(2)}\left(u,\frac{u}{v}\right)&=&\frac{y}{2}
\Biggl[3\pa_{\d}^2D_{1-\d,1,1+\d,1}
-(\ln^2v+\pi^2)~\Phi^{(1)}(v,y)
\Biggr]\, ,\\
\Phi^{(2)}\left(\frac{1}{v},\frac{1}{u}\right)&=&\frac{v}{2}
\Biggl[3\pa_{\d}^2D_{1-\d,1+\d,1,1}-(\ln^2y+\pi^2)~\Phi^{(1)}(v,y)
\Biggr] \, .
\end{eqnarray}

Since our main interest is the function ${ F}^{(2)}(v,Y)$
describing the four-point function of chiral operators at two loops,
we combine the formulae above to get a quantity
\begin{eqnarray}
\nonumber
{\cal S}&=&\frac{1}{u}\Phi^{(2)}\left(\frac{1}{v},\frac{1}{u}\right)+
\Phi^{(2)}\left(\frac{u}{v},u\right)
+\frac{v}{u}\Phi^{(2)}\left(\frac{v}{u},v\right) \\
\label{fin}
&=&\frac{y}{2}\Biggl[ 3\pa_{\d}^2( D_{1-\d,1,1,1+\d}+ D_{1-\d,1,1+\d,1}
+D_{1-\d,1+\d,1,1})\\
\nonumber
&& \hskip 1cm
 -(2\ln ^2v+\ln v\ln y +2\ln^2 y+3\pi^2) \Phi^{(1)}(v,y)\Biggr] \, .
\end{eqnarray}
The remaining step consists in evaluating the Mellin-Barnes integrals for other $D$-functions
involved in (\ref{fin}) with subsequent differentiation of the resulting series.
In this way we arrive at a formula suitable for the study of the OPE of
chiral operators in the direct channel.

It is worth emphasizing  that $\ln^3v$ terms cancel in the final expression for
${\cal S}$. This should not come as a surprise, otherwise one could see the
presence of $\ln^3v$-terms in the four-point function. Such terms would contradict
the general OPE expansion in this order of perturbation theory. Below we
present explicit expressions for the $\ln^2v$ and $\ln v$ terms of the function
${\cal S}$:
\begin{eqnarray}
\label{log2}
{\cal S}_{\ln^2}(v,Y)&=&(1-Y)
\sum_{n,m=0}^{\infty}\frac{Y^m}{m!}\frac{v^n}{(n!)^2}
\frac{\G^2(1+n)\G^2(1+n+m)}{\G(2+2n+m)}\times \\
\nonumber
&\times&
\Biggl(-\frac{1}{4}\ln(1-Y)-\psi(1+n+m)+\psi(2+2n+m) \Biggr)~\ln^2v
\end{eqnarray}
and
\begin{eqnarray}
\nonumber
&&{\cal S}_{\ln}(v,Y)=\frac{1-Y}{2}
\sum_{n,m=0}^{\infty}\frac{Y^m}{m!}\frac{v^n}{(n!)^2}
\frac{\G^2(1+n)\G^2(1+n+m)}{\G(2+2n+m)}\Biggl[
2\ln^2(1-Y) \\
\nonumber
&&+\ln(1-Y)(6\psi(1+n)+5\psi(1+n+m)
-5\psi(2+2n+m))+\pi^2 \\
\nonumber
&&-18\psi^2(1+n)-6\psi^2(2+2n+m)
+12\psi(1+n)(\psi(1+n+m)+\psi(2+2n+m)) \\
\label{Sadd}
&&-9\psi'(1+n)-3\psi'(1+n+m)+6\psi'(2+2n+m)
\Biggr]\ln v
\end{eqnarray}
The non-logarithmic terms are more involved and not essential for our
further study. Substituting (\ref{log2}), (\ref{Sadd}) into (\ref{F2}) and using
representation (\ref{Phi1}) for $\Phi^{(1)}$ we obtain a series representation
for ${ F}^{(2)}$ suitable for the further OPE analysis.

\section{OPE analysis at two loops}
\setcounter{equation}{0}
In this section we employ the expansion
of the functions ${ F}^{(2)}$ found in the previous section to study the operator
algebra of chiral operators at two loops.
Our prime interest will be to confirm the non-renormalization properties
of certain lower-dimensional operators occurring in the operator algebra of chiral operators
as well as to compute the
two-loop anomalous dimensions of some other multiplets.

As was discussed above the non-renormalization property of $O^{\bf 20}_{4,0}$
does not rely on a specific form of the function ${ F}$. However,
non-renormalization of higher-dimensional operators, in particular of descendants
of this operator, can not be unravel without involving  an explicit form of
${ F}$. If we restrict our attention, say, to ${ F}^{(1)}$ or ${ F}^{(2)}$,
then due to the problem of the operator mixing,
an information we get from the OPE analysis is in general
not enough to deduce the individual properties of mixed operators.
However, combining ${ F}^{(1,2)}$ with the knowledge of the $F^{str}$-function
at strong coupling we will be able to trace the perturbative
behavior of some of these operators. This happens due to the fact that
the Yang-Mills multiplets dual to string states become infinitely
massive at strong coupling and do not show up in the corresponding OPE,
whose content is then given by non-renormalized operators and
operators dual to multi-particle gravity states.

According to (\ref{gcpwa}) a CPWA of any tensor contains a multiplier $v^{h/2}$,
where $h$ is treated as an anomalous
dimension. It is then can be decomposed as
\begin{eqnarray}
h=h^{(1)}+h^{(2)}+\ldots \, ,
\end{eqnarray}
where $h^{(1)},~h^{(2)}$ are anomalous dimensions at order $\tilde{\lambda}$, $\tilde{\lambda}^2$
and so on.
Thus, in perturbation theory a term $v^{\frac{1}{2}h}$ is a origin of
logarithmic terms of the form
\begin{eqnarray}
\label{logs}
v^{\frac{1}{2}h}=1+\frac{1}{2} h^{(1)}\ln v+
\left(\frac{1}{2}h^{(2)}\ln v
+\frac{1}{8}(h^{(1)})^2\ln^2v\right)+...
\end{eqnarray}
Here the terms in the brackets occur at order $\tilde{\lambda}^2$ and
should be matched with logarithmic terms in the four-point function
originating from ${ F}^{(2)}$. In particular, the coefficients of the
$\ln v$-terms encode a new information about two-loop anomalous dimensions,
while the ones of $\ln^2v$-terms depend on one-loop anomalous dimensions
having been already found from ${ F}^{(1)}$. Keeping track of the
latter terms is an important consistency a check that perturbative
expansion of the four-point function
fits the corresponding expansion of a sum of CPWAs.

In order not to overload the discussion with formulae, we consider
only the lower-dimensional structure of the OPE for the irreps {\bf 1}, {\bf 20}
and {\bf 105}. Also, we do not write down  the non-logarithmic terms
in the four-point function explicitly but simply present the relevant results wherever
appropriate.
In the following we also assume that for any operator $T$ the ratio
$C^2_{OOT}/C_T$ of the normalization constants occurring in the
corresponding three- and two-point functions with CPOs is kept equal to its free-field value.
A coupling-dependent correction to the constant $C_{OOT}(\tilde{\lambda})$
is introduced in the following way
$$
C_{OOT}(\tilde{\lambda})=C_{OOT}(1+C^{(1)}+C^{(2)}+...),
$$
where $C_{OOT}$ stands for the free-field value and $C^{(i)}$ describes the $i$-th-loop
correction. Below we use the CPWAs normalized as in Section 2 with the exception
of the CPWA of $T_{4,2}$, which we multiply by $-1/4$ to be in agreement with
\cite{AFP2}.

\vskip 0.5cm
{\it Singlet}

The operators of approximate dimension up to 4 emerging in the
singlet projection have already been discussed in \cite{AFP2}.
These are the Konishi scalar ${\cal K}\equiv {\cal K}_{2,0}$, the
Konishi tensor ${\cal K}_{4,2}$, the conserved stress-energy
tensor $T_{\m\n}$, another tensor $\Xi_{4,2}$ which is the lowest component
of a new supersymmetry multiplet and scalar operators of
dimension 4. In particular, with the normalization of
chiral operators we have chosen, the free-field normalization constants are
\begin{eqnarray}
\label{ffnc} \frac{C_{OO{\cal K}}^2}{C_{\cal K}}=\frac{4}{3N^2}\,
, ~~~~ \frac{C_{OO{\cal K}_{4,2}}^2}{C_{{\cal
K}_{4,2}}}=\frac{16}{63N^2}\, ,~~~~
\frac{C_{OO\Xi_{4,2}}^2}{C_{\Xi_{4,2}}}=\frac{16}{35N^2}\, .
\end{eqnarray}

Projecting the two-loop four-point function on the singlet we find
 the following result for the leading terms:
\begin{eqnarray}
-\frac{1}{N^2\, x_{12}^4x_{34}^4}\Biggl[8v+\frac{5}{2}vY
\label{sing}
+\frac{53}{18}vY^2+...\Biggr]\tilde{\lambda}^2\ln v
\, .
\end{eqnarray}
Here the term proportional to $v$ receives a contribution only from
${\cal K}_{2,0}$, for which we have $C^{(1)}_{OO{\cal K
}}=-3\tilde{\lambda}$, $h^{(1)}_{\cal K}=3\tilde{\lambda}$. Thus,
comparison with the corresponding term in the CPWA allows us to
find the two-loop anomalous dimension of the Konishi field
\begin{eqnarray}
h^{(2)}_{\cal K}=-3\tilde{\lambda}^2 \, .
\end{eqnarray}
The two-loop anomalous dimension of the Konishi field has been
previously calculated in \cite{BKRS2} by a different method
and the result obtained there agrees with ours.

The term $vY$ occurs only due to the Konishi field and does not
provide any new information. We therefore consider next the
term $vY^2$, which receives contributions from the Konishi field as
well as from the tensors ${\cal K}_{4,2}$ and $\Xi_{4,2}$. The
contribution of the CPWAs of these fields into the $vY^2\ln v$
term of the four-point amplitude at two loops reads
 \bea
\label{cpc} F(v,Y)|_{vY^2\ln v}&=& \frac{C_{OO{\cal
K}}^2}{18C_{\cal K}}\Big(3C^{(1)}_{OO{\cal K}}h^{(1)}_{\cal
K}+2(h^{(1)}_{\cal
K})^2+3h^{(2)}_{\cal K}\Big)+\\
\nonumber &+& \frac{C_{OO{\cal K}_{4,2}}^2}{8C_{{\cal
K}_{4,2}}}\Big(C^{(1)}_{OO{\cal K}_{4,2}}h^{(1)}_{{\cal
K}_{4,2}}+h^{(2)}_{{\cal K}_{4,2}}\Big) +
\frac{C_{OO{\Xi}_{4,2}}^2}{8C_{{\Xi
}_{4,2}}}\Big(C^{(1)}_{OO{\Xi}_{4,2}}h^{(1)}_{{\Xi}_{4,2}}+h^{(2)}_{{\Xi}_{4,2}}\Big)\,
. \eea Here the free-field normalization constants are given by
(\ref{ffnc}). We also have $h^{(1)}_{\cal K}=h^{(1)}_{{\cal
K}_{4,2}}=3\tilde{\lambda}$, $h^{( 2)}_{\cal K}=h^{(2)}_{{\cal
K}_{4,2}}=-3\tilde{\lambda}^2$,
$h^{(1)}_{{\Xi}_{4,2}}=\frac{25}{6}\tilde{\lambda}$ and
$C^{(1)}_{OO{\cal K }}=-3\tilde{\lambda}$. Using  supersymmetry
we can also determine the correction $C^{(1)}_{OO{\cal
K}_{4,2}}$. Indeed, as was shown in \cite{AES}, supersymmetry
relates the normalizations of the CPWAs corresponding to
different conformal primaries from the same supersymmetry
multiplet. In our case the relevant formula reads (see also
\cite{DOs}) \bea
 \frac{C_{OO{\cal K}_{4,2}}(\tilde\lambda)^2}{C_{{\cal
K}_{4,2}}(\tilde\lambda)} =\frac{(4+h)(6+h)}{6(3+h)(7+h)}
\frac{C_{OO{\cal K}}(\tilde\lambda)^2}{C_{{\cal
K}}(\tilde\lambda)}\, ,\eea where $h$ is the {\it all-loop}
anomalous dimension of the Konishi operator. Expanding this
relation up to the first order in $\tilde{\lambda}$ we find
$C^{(1)}_{OO{\cal K}_{4,2}}=-\frac{89}{28}\tilde{\lambda}$.
Finally, using equation (21) from \cite{AFP2} we obtain the
correction to the three-point function involving CPOs and the tensor
$\Xi_{4,2}$,
$C^{(1)}_{OO{\Xi}_{4,2}}=-\frac{1025}{252}\tilde{\lambda}$. We now
substitute all these findings in (\ref{cpc}) and match the latter
to $-\frac{53}{18}\frac{\tilde{\lambda}^2}{N^2}$, which is the
coefficient of $vY^2\ln v$ term in (\ref{sing}). In this way we
find the anomalous dimension of $\Xi$ at two loops:\footnote{In
the first version of this paper an erroneous value for
$C^{(1)}_{OO{\cal K}_{4,2}}$ has been used which led to an 
incorrect result for the two-loop anomalous dimension of $\Xi$.
We are grateful to Francis Dolan and Hugh Osborn for pointing out
a possible error in our original calculation.}
\begin{eqnarray}
h^{(2)}_{\Xi}=-\frac{925}{216}\tilde{\lambda}^2.
\end{eqnarray}
Similarly  to the dimension of the Konishi field, the anomalous
dimension of $\Xi$ is negative. Quite remarkably, this value
coincides  with the one obtained in \cite{KL1,KL2} by using a
different approach based on the DGLAP and BFKL equations. In fact, in
\cite{KL1,KL2} the two-loop (and even the three-loop) anomalous dimensions for the
conformal twist two operators of arbitrary spin have been found.
We note that our calculation can be generalized to twist two
operators of spin ${l}$ by analyzing the coefficient of
$vY^{l}\ln v$ in the CPWA expansion (of the singlet channel) of
the four-point amplitude at two loops.

In addition to the operators discussed above, the terms $v^2$ and $v^2Y$ not indicated
in (\ref{sing}) receive contributions from the scalar operators of dimension 4.
In \cite{AFP2} we assumed that the free-field double trace operator $O^{\bf 1}$ undergoes
a splitting into a sum of operators $O_i$. However, despite having at
our disposal the result for
the two-loop four-point function, the relatively large number $(\geq 3)$
of mixed operators does not allow us to find their individual
anomalous dimensions and free-field normalization constants.

Finally, analyzing the leading non-logarithmic terms in the four-point function
one obtains the following results for the two-loop corrections to the ratio
of the normalization constants for ${\cal K}$:
\begin{eqnarray}
C^{(2)}_{\cal K}=\frac{3}{2}(7+3\zeta(3))\frac{\tilde{\lambda}^2}{N^2}\, .
\end{eqnarray}

\vskip 0.5cm
{\it Irrep {\bf 20}}

Here we show that the content of the operator algebra formed by operators
up to approximate dimension 6 and
transforming in the irrep $\bf 20$ of $SO(6)$
can be depicted as follows:
\begin{center}
\nonumber
\begin{tabular}{c|ccccc}
$\D$ & spin &                          & & & \\
\hline
  &              &                     & & & \\
2 & $O_{2,0}$          &                     & & & \\
4 & $O_{4,0}$ & ${\cal K}_{4,2}$ & & & \\
6 &  ${O}_{6,0}, {\cal K}_{6,0}, \Xi_{6,0} $     & $T_{6,2}$  & $\Xi_{6,4}$ & &
\end{tabular}
\end{center}

The operator $O_{2,0}$ is the CPO and $O_{4,0}$ is an operator
whose non-renormalization
property was discussed in Section 2. Below we demonstrate
that in addition to these operators
there exist a scalar $O_{6,0}$ with vanishing anomalous dimension.
We will also see that
a free-field scalar $T_{6,0}$ splits in perturbation
theory into the sum of three operators
belonging to different representations of supersymmetry.

As was already shown in \cite{AFP2} ({\it c.f.} Section 2)
the lowest-dimensional operator in irrep {\bf 20}
that receives anomalous dimension is the
second rank tensor Konishi tensor ${\cal K}_{4,2}$ with the free-field ratio
$C_{OO{\cal K}_{4,2}}^2/C_{{\cal K}_{4,2}}=\frac{80}{9N^2}$.
Extending the free-field and the one-loop analysis of \cite{AFP2}
to dimension 6 operators, it is not difficult to show that a tensor $T_{6,4}$
has the one-loop anomalous dimension $h^{(1)}=\frac{25}{6}\tilde{\lambda}$, i.e. it is
the same as for the tensor $\Xi_{4,2}$ occurring in the singlet projection.
Thus,  $T_{6,4}\equiv\Xi_{6,4}$
belongs to the $\Xi$-multiplet. With our convention for normalization of CPWAs
its free-field ratio of the normalization constants is
\begin{eqnarray}
\frac{C_{OO\Xi_{6,4}}^2}{C_{\Xi_{6,4}}}=\frac{4}{21N^2}\, .
\end{eqnarray}

To proceed it is useful to recall the strong coupling
result \cite{AFP1} for the four-point function of chiral operators
projected onto {\bf 20}.
For the first few leading terms we get
\begin{eqnarray}
\label{p20ads}
\frac{1}{N^2\, x_{12}^4x_{34}^4} \Biggl[
\frac{40}{3}v F_1(Y)
+v^2\left(\frac{26}{9}+\frac{26}{9}Y+\frac{119}{45}Y^2\right)
+\frac{2}{15}v^3
 -\frac{4}{3}v^2\ln v \left(Y^2-v-\frac{3}{2}vY\right)
\Biggr] \, ,
\end{eqnarray}
where we have written out explicitly both logarithmic and non-logarithmic
terms. Here a function $F_1(Y)=-Y^{-1}\ln(1-Y)$ provides a complete
$Y$-contribution of the CPWA of a dimension 2 scalar that is the chiral
operator $O_{2,0}^{\bf 20}$ itself. Such a structure of the $v$-term allows one
to conclude that all ``single-trace'' rank-$l$ operators of dimension $2+l$
decouple  at strong coupling, $\Xi_{6,4}$ among them \cite{AFP1}. {}From
(\ref{p20ads}) one may see that the coefficient of the $\ln v$-term matches
exactly the leading terms of the CPWA of a tensor $T_{6,2}$, in particular,
this coefficient does not receive contribution from the CPWA of a scalar
$T_{6,0}$. Thus, we have two options: either $T_{6,0}$ is non-renormalized or
it is absent in the strong coupling OPE. Let us show that the first option is
realized. To this end we study the non-logarithmic terms in (\ref{p20ads}).

We represent the $1/N^2$ corrections
to a normalization constant in the usual way as e.g.
$
C_{\D\, ,l}=C_{\D\, ,l}\left(1+C^{(1)}_{\D\, ,l}\right) \, ,
$
where $C_{\D\, ,l}$ on the r.h.s. is a leading term in $1/N^2$ and $C^{(1)}_{\D\, ,l}$ is
a next $1/N^2$-correction. In particular, the $v^2Y^2$-term contains
the contribution from CPWAs of
$O_{2,0}$, $O_{4,2}$, and of $T_{6,2}$.
The contribution of the CPWA of $T_{6,0}$ is absent since it starts from $v^3$.
Thus, matching the $v^2Y^2$-terms we find
\begin{eqnarray}
\label{C62}
\frac{C_{OOT_{6,2}}^2}{C_{6,2}} C_{6,2}^{(1)}=\frac{2}{45N^2} \, .
\end{eqnarray}
In a similar way studying the contribution of CPWA's to the $v^3$-term in (\ref{p20ads})
and taking into account (\ref{C62}) we find
\begin{eqnarray}
\frac{C_{OOT_{6,0}}^2}{C_{6,0}} C_{6,0}^{(1)}=-\frac{4}{9N^2} \, .
\end{eqnarray}
Thus, we clearly see that scalar $T_{6,0}$ is present in the strong coupling OPE
but does not receive any anomalous dimension.

To get more insight we consider the projection of the free-field four-point function
onto irrep. {\bf 20}:
\begin{eqnarray}
\label{p20free}
\frac{1}{\, x_{12}^4x_{34}^4} \Biggl[
\frac{40}{3N^2}v F_1(Y)+
v^2\left(2+\frac{2}{3N^2}+\left(2+\frac{2}{3N^2}\right)Y
+\left(3+\frac{2}{3N^2}\right)Y^2 +... \right)
\Biggr] \, .
\end{eqnarray}
Note that the higher $v$-terms are absent.
The terms $v^2Y^2$ and $v^3$ get contributions from
the CPWAs of the CPO, ${\cal K}_{4,2}$, $T_{6,2}$, $O_{4,2}$ and $\Xi_{6,4}$
and this allows us to find the free-field values of the normalization constants
\begin{eqnarray}
\label{coupl6}
\frac{C_{OOT_{6,2}}^2}{C_{6,2}}=\frac{6}{5}+\frac{1}{15N^2},~~~~~~~
\frac{C_{OOT_{6,0}}^2}{C_{6,0}}=\frac{2}{3}-\frac{1}{9N^2}\, .
\end{eqnarray}
Thus, we see that $1/N^2$ strong coupling corrections
to the constants of $T_{6,0}$ and $T_{6,2}$ do not coincide
with their free-field counterparts. This means that operators
$T_{6,0}$ and $T_{6,2}$ undergo a splitting at weak coupling
into a sum of operators with different perturbative behavior of anomalous dimensions.
In particular the operator  $T_{6,0}$ should contain in the split
a non-renormalized operator.

Extension of the one-loop analysis performed in \cite{AFP2} to the operators of
dimension 6 allows us to establish the following relations
\begin{eqnarray}
\label{S1}
\sum_i \frac{C_{OOT_{6,0}^i}^2}{C_{6,0}^i}(h^i_{6,0})^{(1)}=\frac{11}{9}
\frac{\tilde{\lambda}}{N^2}\, ,~~~~
\sum_i \frac{C_{OOT_{6,2}^i}^2}{C_{6,2}^i}(h^i_{6,2})^{(1)}=\frac{5}{9}
\frac{\tilde{\lambda}}{N^2} \, ,
\end{eqnarray}
where we have taken into account that above discussed operators
split at one-loop.

Finally we can use the whole power of our formulae to extract the one-loop anomalous
dimensions by looking at the $\ln^2v$ terms
in the two-loop four-point function
projected on the irrep {\bf 20}. For the first few leading terms we find
\begin{eqnarray}
\label{p20}
\frac{1}{N^2}\Biggl[\frac{5}{2}v(Y^2-v-vY)
-v^2\left(\frac{5}{2}+\frac{5}{2}Y+\frac{205}{108}Y^2\right)
-\frac{73}{108}v^3
\Bigg]\tilde{\lambda}^2\ln^2{v}\, ,
\end{eqnarray}
where the first term is distinguished to emphasize the contribution
of the CPWA of the tensor ${\cal K}_{4,2}$. The essence of our analysis are
the following equations:
\begin{eqnarray}
\label{S2}
\sum_i \frac{C_{OOT_{6,0}^i}^2}{C_{6,0}^i}[(h^i_{6,0})^{(1)}]^2=\frac{124}{27}
\frac{\tilde{\lambda}^2}{N^2}\, , ~~~~
\sum_i \frac{C_{OOT_{6,2}^i}^2}{C_{6,2}^i}[(h^i_{6,2})^{(1)}]^2=\frac{205}{27}
\frac{\tilde{\lambda}^2}{N^2} \, .
\end{eqnarray}

Consider now $T_{6,0}$ and make {\it an assumption}
that in perturbation theory this operator splits into  three operators,
one $O_{6,0}$ is non-renormalized, the second, ${\cal K}_{6,0}$,
is from the Konishi multiplet and the third one, $\Xi_{6,0}$, is an operator
whose anomalous dimension that we are going to find\footnote{We denote
this operator by $\Xi$ since as will become clear in a moment
it belongs to the $\Xi$-multiplet.}.
The free-field normalization constant corresponding to   ${O}_{6,0}$ should be the same
as we have found from the strong coupling result, i.e.
\begin{eqnarray}
\frac{C_{OO{ O}_{6,0}}^2}{C_{O_{6,0}}}=\frac{2}{3}-\frac{4}{9N^2}.
\end{eqnarray}
Subtracting it from the free-field result (\ref{coupl6}) we are left with
the sum of the constants of the operators ${\cal K}_{6,0}$ and $\Xi_{6,0}$:
\begin{eqnarray}
\frac{C_{OO{\cal K}_{6,0}}^2}{C_{{\cal K}_{6,0}}}
+\frac{C_{OO\Xi_{6,0}}^2}{C_{\Xi_{6,0}}}=\frac{1}{3N^2}\, .
\end{eqnarray}
This equation together with (\ref{S1}) and (\ref{S2}) provides
a system of three equations for three unknown variables that are
normalization constants and the anomalous dimension of $\Xi_{6,0}$. Solving the system we obtain
\begin{eqnarray}
\frac{C_{OO{\cal K}_{6,0}}^2}{C_{{\cal K}_{6,0}}}=\frac{1}{7N^2} \, ,~~~~~
\frac{C_{OO\Xi_{6,0}}^2}{C_{\Xi_{6,0}}}=
\frac{4}{21 N^2} \, ,~~~~~
h_{\Xi_{6,0}}^{(1)}=\frac{25}{6}\tilde{\lambda}
\end{eqnarray}
which clearly shows that $\Xi_{6,0}$ belongs to the $\Xi$-multiplet.
As to $T_{6,2}$, the corresponding analysis is complicated
by the fact that this operator(s) is present at strong coupling
with a finite anomalous dimension and the information we can extract
from the weak/strong four-point functions
is not enough to establish its split components.

\vskip 0.5cm
{\it Irrep {\bf 105}}

As was shown in \cite{AFP1} the rank-$2k$ tensors $O_{4+2k,2k}$
and $O_{6+2k,2k}$ transforming in the irrep {\bf 105} are non-renormalized
in the strong coupling limit. As we have seen in Section 2
the non-renormalization property of $O_{4+2k,2k}$ is a (non-perturbative)
consequence of the non-renormalization theorem of \cite{EPSS}.
The strong coupling behavior
of the normalization constant of  $O_{6+2k,2k}$ indicates however
that a free-field theory operator $T_{6+2k,2k}$
splits is perturbation theory into a sum of operators, therefore the
unraveling of its non-renormalized component  $O_{6+2k,2k}$
requires the explicit knowledge of the function $F(v,Y)$.

Here, restricting our attention
to the dimension 6 operators and assuming that there exist
$O_{6,0}$ and $O_{6,2}$ that are non-renormalized,
we reveal the corresponding weak coupling content of the operator algebra.
The subsequent treatment does not involve the knowledge of the two-loop
four-point function and it relies only on the free-field, the one loop and the strong
coupling considerations.

Analyzing the free-field four-point function we find
the free-field couplings
\begin{eqnarray}
\label{frc}
\frac{C_{OOO_{4,0}}^2}{C_{4,0}}=2+\frac{4}{N^2}\, , ~~~
\frac{C_{OOT_{6,2}}^2}{C_{6,2}}=\frac{6}{5}+\frac{2}{5N^2}\, ,~~~
\sum_i\frac{C_{OOT_{6,0}^i}^2}{C_{6,0}^i}=\frac{2}{3}-\frac{2}{3N^2}\, .
\end{eqnarray}
Here $O_{4,0}$ is an operator belonging to the short multiplet
whose non-renormalization property is well-known.
For $T_{6,0}$ we assume a perturbative splitting.

At strong coupling we find however a non-renormalized  operator $O_{6,0}$ with the $1/N^2$
correction to the normalization constants: $-2/N^2$.
Thus, $C_{OOO_{6,0}}^2/C_{O_{6,0}}=\frac{2}{3}-\frac{2}{N^2}$,
i.e. it is different from (\ref{frc}). We assume that this difference
is due to the fact that $T_{6,0}$ splits in perturbation theory into a sum of {\it two}
operators: $O_{6,0}$ and another operator ${\cal K}_{6,0}$ with a free-field value
of the ratio $C_{OO{\cal K}_{6,0}}^2/C_{{\cal K}_{6,0}}=\frac{4}{3N^2}$.
With this assumption we can now analyze the one-loop four-point function and
determine the one-loop anomalous dimension of ${\cal K}_{6,0}$ that turns out
to be $h^{(1)}_{{\cal K}_{6,0}}=3\tilde{\lambda}$. Thus, ${\cal K}_{6,0}$
belongs to the Konishi multiplet.

\vskip 0.3cm {\bf Acknowledgements}
\small

\noindent G.A. was supported by the DFG and by the European Commission RTN
programme HPRN-CT-2000-00131 in which G.A. is associated to U. Bonn, and in
part by RFBI grant N99-01-00166 and by INTAS-99-1782. G.A. is grateful to S.
Frolov, S. Kuzenko and to S. Theisen for useful discussions. A.C.P. wishes to
thank H. Osborn for valuable discussions. He is partially supported by the
European Commission RTN program HPRN-CT-2000-00131 in which he is associated to
the University of Torino. B.E. and E.S. are grateful to I. Todorov, D.Z.
Freedman and E. D'Hoker, and E.S. to V. Dobrev, S. Ferrara and V. Petkova for
stimulating discussions.


\normalsize

\section*{Appendix A \\ Truncation property of the CPWA}
\setcounter{equation}{0}
The CPWA for the exchange of a tensor arbitrary non-integer
dimension $\Delta$ and spin $l$ between two pairs of scalar fields
was calculated in \cite{HPR}. We state their result for the special
case of space-time dimension $d=4$ and dimension of the outer
scalar operators $\tilde\Delta=2$. The overall normalization
factor $\beta_{\tilde\Delta;\Delta,l}$ is omitted since the
$\Gamma$-functions in it cancel in this case.
\begin{eqnarray}
{\cal H}_{\Delta,l} = \frac{1}{(x_{12}^2 \, x_{34}^2)^{2}} \,
\sum_{n,m=0}^\infty \frac{v^n \, Y^m}{n! \, m!} \sum_{M=0}^l
\frac{c_l^{(M)}}{2^M \, c_l^{(l)}} \sum_{n_{i}=0}^M (-1)^{n_1 +
n_3} \frac{M!}{n_1! \, n_2! \, n_3! \, n_4!} \, (1-Y)^{n_2} \times
\nonumber \\ v^{\frac{1}{2}(\Delta-M)} \, \alpha(\delta_2)
\alpha(\delta_4) \alpha(\Delta) \frac{(\delta_1)_n \,
(2-\delta_2)_n \, (\delta_3)_{n+m} \, (2
-\delta_4)_{n+m}}{(\Delta)_{2n+m} \, (\Delta-\frac{1}{2}d+1)_n} \,
\label{cpwa}
\end{eqnarray}
Here
\begin{equation}
\alpha(x) \, =  \, \frac{\Gamma(2-x)}{\Gamma(x)} \, ,
\end{equation}
\begin{eqnarray}
\delta_1 & = \frac{1}{2}(\Delta-M) + n_4 + n_1, \; \; \; \; \; \;
\delta_2 & = 2 - \frac{1}{2}(\Delta+M) + n_1 + n_2, \nonumber
\\ \delta_3 & = \frac{1}{2}(\Delta-M) + n_2 + n_3, \; \; \; \;
\; \; \delta_4 & = 2 - \frac{1}{2}(\Delta+M) + n_3 + n_4
\end{eqnarray}
and the summation over the $n_{i}$ is such that $\sum n_{i} \, =
\, M$.

We split the dimension of the exchanged operator as
$\Delta=\Delta_0-h$, where $\Delta_0$ is an integer and $-1 \leq h
< 1 $ is the anomalous part of the dimension. The overall factor
$v^{h/2}$ may be pulled out and is ignored in the following.

We set out to show that the lowest terms in the $v,Y$ expansion of
(\ref{cpwa}) are of the form $v^{\frac{1}{2}(\Delta_0-k)} \, Y^k$,
where $k=l, \, l-2, \, l-4, \, \ldots,$ but $k \geq 0$ . This
requires proving the cancellation of some powers of $Y$ arising
from $(1-Y)^{n_2}$. It suffices to consider each value of $n,m,M$
separately, hence we can restrict our attention to the sum over
$n_{i}$. There are three summations because $\sum n_{i} \, = \,
M$. Define $N_4 = n_3+n_4$. A substantial simplification is
obtained by rewriting
\begin{equation}
\sum_{n_{i}=0}^M \, = \, \sum_{N_4=0}^M \, \sum_{n_1=0}^{M-N_4} \,
\sum_{n_4=0}^{N_4}
\end{equation}
since then $\delta_2,\delta_4$ depend only on $M,N_4$ but not on
the remaining two counters $n_1,n_4$.

The proof can in fact be established for fixed $N_4$. Consider
\begin{equation}
S = \sum_{n_1=0}^{M-N_4} \, \sum_{n_4=0}^{N_4} \, (-1)^{n_1+n_3}
\frac{M!}{n_1! \, n_2! \, n_3! \, n_4!} \, (1-Y)^{n_2}
(\delta_1)_n \, (\delta_3)_{n+m} \, ,
\end{equation}
the other terms being constant for fixed $n,m,M,N_4$. Rearrange as
\begin{eqnarray}
S & = & (-1)^{N_4} \begin{pmatrix} M \\ M-N_4 \end{pmatrix} \,
\sum_{n_1=0}^{M-N_4} (-1)^{n_1}
\begin{pmatrix} M-N_4 \\ n_1 \end{pmatrix} \, \sum_{n_4=0}^{N_4}
(-1)^{n_4} \begin{pmatrix} N_4 \\ n_4 \end{pmatrix} \nonumber
\\ & \times & (1-Y)^{M-N_4-n_1} (\delta_1)_n \, (\delta_3)_{n+m}
\\ & = & (-1)^{N_4} \begin{pmatrix} M \\ M-N_4
\end{pmatrix} \sum_{k=0}^{M-N_4} (-Y)^k \sum_{n_1=0}^{M-N_4-k}
(-1)^{n_1} \begin{pmatrix} M-N_4 \\ n_1 \end{pmatrix} \,
\begin{pmatrix} M-N_4-n_1 \\ k \end{pmatrix} \nonumber \\  &
\times & \sum_{n_4=0}^{N_4} (-1)^{n_4} \begin{pmatrix} N_4
\\ n_4 \end{pmatrix} (\delta_1)_n \, (\delta_3)_{n+m} \\& = &
(-1)^{N_4} \begin{pmatrix} M \\ M-N_4 \end{pmatrix}
\sum_{k=0}^{M-N_4} (-Y)^k \begin{pmatrix} M-N_4 \\ k \end{pmatrix}
\nonumber \\ & \times & \Biggl[ \sum_{n_1=0}^{M-N_4-k} (-1)^{n_1}
\begin{pmatrix} M-N_4-k \\ n_1 \end{pmatrix} \,
\sum_{n_4=0}^{N_4} (-1)^{n_4} \begin{pmatrix} N_4 \\ n_4
\end{pmatrix} (\delta_1)_n \, (\delta_3)_{n+m} \Biggr] \, .
\end{eqnarray}
All terms in the last line apart from the Binomial coefficients
depend only on the sum $N_1=n_4+n_1$. Using
\begin{equation}
\sum_{n+m=p} \begin{pmatrix} N \\ n \end{pmatrix} \,
\begin{pmatrix} M \\ m \end{pmatrix} \, = \, \begin{pmatrix} M+N
\\ p \end{pmatrix}
\end{equation}
we replace the double sum in the square bracket by
\begin{equation}
\sum_{p=0}^{M-k} (-1)^p
\begin{pmatrix} M-k \\ p \end{pmatrix} \biggl(\frac{1}{2}(\Delta-M)+p \biggr)_n
\, \biggl([\frac{1}{2}(\Delta-M)+k] + [(M-k)-p] \biggr)_{n+m} \, .
\label{magic}
\end{equation}
It will be demonstrated below that this sum vanishes if $M-k>2n+m$
and that is equals $(-1)^n \, (2n+m)! \,$ if $M-k=2n+m$. The
lowest power of $Y$ occurring in $S$ is therefore $Y^{(M-2n-m)}$
if $M \geq 2n+m$ and $Y^0$ if $M < 2n+m$.

Recall that the complete expression for the CPWA (\ref{cpwa})
includes
\begin{equation}
v^{(\frac{h}{2})} \, v^{(\frac{1}{2}(\Delta_0-M)+n)} Y^m
\end{equation}
the product of which with $Y^{(M-2n-m)}+\ldots$ yields
\begin{equation}
v^{(\frac{h}{2})} \, v^{(\frac{1}{2}(\Delta_0-(M-2n))}
(Y^{(M-2n)}+\ldots)
\end{equation}
so that the lowest term is in fact of the type postulated above,
since the Gegenbauer coefficients $c_l^{(M)}$ are non-vanishing
only if $M,l$ are both odd or both even.

Referring to the ``total power" $T(v^{(\frac{h}{2})} v^n Y^m)
\equiv 2n+m$ introduced above we find only terms with $T \geq
\Delta_0$.

Last, if $M<2n+m$ we find
\begin{equation}
v^{(\frac{h}{2})} \, v^{(\frac{1}{2}(\Delta_0-M)+n)} Y^m
(Y^0+\ldots) \, .
\end{equation}
The lowest of these terms comes with $T = \Delta_0 -M + 2n + m >
\Delta_0$; they are never leading.

It remains to prove the vanishing of (\ref{magic}) for $M-k>2n+m$.
One may check by explicit calculation that
\begin{eqnarray}
&& (X+p)_a \, (Y+P-p)_b  - (X+(p+1))_a \, (Y+P-(p+1))_b = \\ && b
\, (X+p)_a \, ((Y+1)+(P-1)-p)_{b-1} - a \, ((X+1)+p)_{a-1} \,
(Y+(P-1)-p)_b \nonumber
\end{eqnarray}
Using Pascal's triangle:
\begin{eqnarray}
&& \sum_{p=0}^P (-1)^p
\begin{pmatrix} P \\ p \end{pmatrix} (X+p)_a (Y+P-p)_b =
\sum_{p=0}^{P-1} (-1)^p \begin{pmatrix} P-1 \\ p \end{pmatrix} \\
&& \times \biggl( b \,(X+p)_a \, ((Y+1)+(P-1)-p)_{b-1} - a \,
((X+1)+p)_{a-1} \, (Y+(P-1)-p)_b \biggr) \, . \nonumber
\end{eqnarray}
On iterating this step the sum vanishes if $P>a+b$ and is equal to
$(-1)^a P\,!\,$ if $P=a+b$.


\section*{Appendix B \\ Analytic continuation of $\Phi(v,y|\d)$}
\setcounter{equation}{0}
Here we discuss the problem of the analytic continuation of the function
$\Phi(v,y|\d)$ to the conformal variables describing the crossed channels.

Recall the Mellin-Barnes representation for the
 $D_{\D_1\D_2\D_3\D_4}$-functions \cite{AFP1}:
\begin{eqnarray}
&&D_{\D_1\D_2\D_3\D_4}(v,y)=\int \frac{d\l ds}{(2\pi i)^2}\Biggl[
\G(-\l)\G(-s)\G(\frac{\D_1+\D_2-\D_3-\D_4}{2}-\l) \\
\nonumber
&&\hspace{-0.3cm}\times
\G(\frac{\D_1+\D_3-\D_2-\D_4}{2}-s)
\G(\frac{\D_2+\D_3+\D_4-\D_1}{2}+s+\l)\G(\D_4+s+\l)v^{\l}\left(\frac{v}{u}\right)^s \Biggr] \, .
\end{eqnarray}
Comparing this formula with (\ref{PHId}) one obtains (\ref{A}).
On the other hand, the $D$-function has an integral representation
in terms of Schwinger parameters (see e.g. \cite{AFP1}),
so that for the case under consideration one gets
\begin{eqnarray}
\nonumber
\Phi\left(v,\frac{v}{u}|\d\right)=2\int dt_1dt_2dt_3dt_4~ t_1^{-\d}t_4^{\d}
exp\Biggl[-t_1t_2-t_1t_3-t_1t_4-t_2t_3-\frac{v}{u}t_2t_4-vt_3t_4  \Biggr]\, .
\end{eqnarray}
{}From here we immediately read off a representation for $\Phi$ in a crossed
channel, e.g.,
\begin{eqnarray}
\nonumber
\Phi\left(u,\frac{u}{v}|\d\right)=2\int dt_1dt_2dt_3dt_4~ t_1^{-\d}t_4^{\d}
exp\Biggl[-t_1t_2-t_1t_3-t_1t_4-t_2t_3-ut_2t_4-\frac{u}{v}t_3t_4  \Biggr]\, .
\end{eqnarray}
Note that under the following rescaling of integration variables
\begin{eqnarray}
t_1\to \l t_1, ~~~t_2\to \frac{1}{\l} t_2, ~~~t_3\to \frac{1}{\l} t_3,
~~~t_4\to \frac{1}{\l} t_4 \,
\end{eqnarray}
the integral takes the form
\begin{eqnarray}
\nonumber
\Phi\left(u,\frac{u}{v}|\d\right)&=&\frac{2}{(\l^2)^{1+\d}}\int dt_1dt_2dt_3dt_4
~ t_1^{-\d}t_4^{\d}
\\\nonumber
&\times&
exp\Biggl[-t_1t_2-t_1t_3-t_1t_4-\frac{1}{\l^2}t_2t_3
-\frac{u}{\l^2}t_2t_4-\frac{u}{\l^2 v}t_3t_4  \Biggr]\, .
\end{eqnarray}
Now we may choose $\l^2=\frac{u}{v}$ and perform the change of variables
$t_2\leftrightarrow t_3$ and then $t_3\leftrightarrow t_4$.
Finally, by using the Mellin-Barnes representation for the integral on the r.h.s.,
we arrive at the first formula in (\ref{ac}).
The second formula in (\ref{ac}) is proved in an analogous manner.

The functions $D_{1-\d,1,1+\d,1}$ and $D_{1-\d,1+\d,1,1}$
have the following Mellin-Barnes representation
\begin{eqnarray}
\nonumber
D_{1-\d,1,1+\d,1}(v,Y)&=&v^{-\d/2}
\int \frac{d\l ds}{(2\pi i)^2}\G^2(-s)
\G(-\l+\d/2)\G(-\l-\d/2)\G(1+\l+s+\d/2)
\\
\label{MB2}
&\times &\G(1+\l+s-\d/2)v^{\l}y^{s} \, \\
\nonumber
D_{1-\d,1+\d,1,1}(v,Y)&=&y^{-\d/2}
\int \frac{d\l ds}{(2\pi i)^2}\G^2(-\l)
\G(-s+\d/2)\G(-s-\d/2)\G(1+\l+s+\d/2)
\\
\label{MB3}
&\times &\G(1+\l+s-\d/2)v^{\l}y^{s} \,
\end{eqnarray}

\end{document}